# ANISOTROPIC COSMOLOGICAL MODEL WITH VARIABLE $G$ and $\Lambda$

S. K.Tripathy[1,2†], D.Behera[1*] and T.R.Routray[2#]

1. Department of Physics, Government College of Engineering Kalahandi, Bhawanipatna, Odisha-766001(India)   2. School of Physics, Sambalpur University , Jyotivihar, Sambalpur, Odisha-768019 (India)

E-mail:† tripathy_sunil@rediffmail.com, * dipadolly@rediffmail.com, # trr1@rediffmail.com

**Abstract**

Anisotropic Bianchi-III cosmological model is investigated with variable gravitational and cosmological constants in the framework of Einstein's general relativity. The shear scalar is considered to be proportional to the expansion scalar. The dynamics of the anisotropic universe with variable $G$ and $\Lambda$ are discussed. Without assuming any specific forms for $\Lambda$ and the metric potentials, we have tried to extract the time variation of $G$ and $\Lambda$ from the anisotropic model. The extracted $G$ and $\Lambda$ are in conformity with the present day observations. Basing upon the observational limits, the behavior and range of the effective equation of state parameter are discussed.

Cosmological models with variable Newtonian gravitational constant $G$ and cosmological constant $\Lambda$ are of great interest in recent times. After Dirac's proposal for the variation of the so called fundamental physical constants, from his Large Number Hypothesis [1,2], there have been many attempts to modify Einstein's general relativity with the incorporation of time varying Newtonian Gravitational constant $G$. Brans and Dicke [2] followed Mach's principle and suggested that the gravitational field is mediated by a long range scalar field $\phi$ that has an inverse dimension of $G$. In the Hoyle-Narlikar steady state model, in Machian sense, the inertia of a particle originates from the rest of the matter present in the universe which leads to a variable $G$ [4]. Empirically, local constraints on the rate of variation of $G$ can be derived from Lunar Laser Ranging. From the ranging data to the Viking landers on Mars, the simple time variation in the effective Newtonian gravitational constant $\frac{\dot{G}}{G}$ could be limited to $(0.2 \pm 0.4) \times 10^{-11} \ yr^{-1}$ [5]. However, the quoted errors in this estimation are much large, which may have stemmed from the lack of knowledge of the masses of asteroids. There are also measurements which offer a variation of such fundamental constants. The experimental range of variation of $\frac{\dot{G}}{G}$ seems to be $10^{-12} \ yr^{-1} < \frac{\dot{G}}{G} < 10^{-10} \ yr^{-1}$. For some recent discussions, based on white dwarf asteroseismology, one may refer to Ref.[6]. Recently, many authors such as Arbab [7- 9], Singh et al. [10], Tiwari [11], Mukhopadhyay et al.[12], Behera et al. [13] , Ram and Verma [14], Singh [15] have discussed $G$ varying cosmologies.

The analysis of the Hubble diagram of distant type Ia Supernovae provides strong evidence that the universe is currently undergoing an accelerated phase of expansion [16-19]. These observations strongly suggest that the content of the universe is largely of non-baryonic origin usually referred to as dark matter. The baryonic content of the universe is only about 5%. In classical Friedman models with zero curvature, the accelerated phase of expansion of the universe is generally explained by a cosmological constant. A non-zero positive cosmological constant $\Lambda \sim 2h^2 \times 10^{-56} cm^{-2}$ is required to account for the observations [16-19]. The cosmological constant problem, an old and yet unsettled problem in physics [20-23], centers around the large difference in the theoretical and observational values of the cosmological constant. The present small but non-zero cosmological constant requires a fine tuning problem in physics. In order to bridge the gap, an evolving cosmological constant has been considered by many authors [24-27]. Berman [28, 29], Bertolami [24, 25], Endo and Fukui [30] and Canuto et al. [31] have suggested an inverse square time varying form of the cosmological constant i.e. $\Lambda \sim t^{-2}$ where as Arbab in his work [8] used $\Lambda = 3\beta H^2$, where $\beta$ is a constant and $H$ is the Hubble parameter. In some recent works Borges and Canerio [32] and Tiwari [11] have considered a cosmological term proportional to the Hubble parameter. Singh [33] found that $\Lambda$ does not necessarily vary in inverse square manner with time as occurring in flat models. In a recent work [13], we have shown that, the time variation of $G$ is intimately related to the time variation of the cosmological constant $\Lambda$. If one considers a time varying cosmological constant term, then it is necessary that one should consider a time varying gravitational constant term in the usual Einstein field equation. This approach is interesting in the sense that it provides a link in the variation the Newtonian gravitational constant with the cosmological constant while leaving the form of Einstein field equations formally unchanged.

In the present work, we have investigated Bianchi type –III (BIII) bulk viscous anisotropic cosmological model in the framework of Einstein's relativity with variable gravitational and cosmological constants. The motivation behind the present work is to extract the time variation of the Newtonian gravitational constant and the cosmological constant from the dynamics of the model investigated keeping an eye on the recent observations related to the accelerated expansion of the universe. Basing upon the observational limits, we have tried to focus on the behavior and range of the effective equation of state parameter. In order to get viable models, we have assumed that the shear scalar is directly proportional to the scalar expansion and the contribution of the bulk viscosity to the total pressure is proportional to the rest energy density. Such a barotropic bulk viscous pressure is required to explain the accelerated expansion of the universe [34-35].

The accelerated phase of expansion of the universe is usually modeled through a cosmological constant in the usual classical Einstein's field equation

$$R_{ij} - \frac{1}{2}g_{ij}R = -8\pi G(t)T_{ij} + \Lambda(t)g_{ij} \qquad (1)$$

where the symbols have their usual meaning. The cosmological constant is considered to be time dependent in order to account for the fine tuning problem arising in the so called cosmological constant problem in physics. The time variation of Newtonian constant is intimately related to the time variation of cosmological constant in classical models [13] and hence one has to take simultaneous time variation of $G$ and $\Lambda$ in the classical field equations. In the field equation (1), the unit is chosen in such a manner that the speed of light in vacuum is unity. The universe is usually described by a perfect fluid distribution. In order to take into account the dissipative phenomenal occurring in the cosmic fluid, we consider that the contribution to the energy-momentum tensor should come from bulk viscosity in addition to the usual fluid. The presence of bulk viscosity in the cosmic fluid has already been recognized in connection with the observed accelerated expansion [34-40].

In commoving coordinates, the energy-momentum tensor is given by

$$T_{ij} = (\rho + \bar{p})u^i u^j + \bar{p} g_{ij} \qquad (2)$$

where $u^i$ are the four velocity vectors defined as $u^i = \delta_4^i$ and they satisfy the relation $u^i u^j = -1$. $\rho$ is the proper rest energy density and $\bar{p}$ is the total effective pressure which includes the proper pressure $p$ and contribution of bulk viscosity to pressure such that $\bar{p} = p - \xi(t) u^l_{;l}$, $\xi(t)$ being the time dependent bulk viscous coefficient.

Anisotropic BIII universe is modeled through the metric

$$ds^2 = -dt^2 + A^2 \, dx^2 + B^2 \, e^{-2hx} \, dy^2 + C^2 dz^2 \qquad (3)$$

where the metric potentials $A, B$ and $C$ are considered as the functions of cosmic time only. The expansion scalar $\theta = u^l_{;l}$ for this metric can be expressed as

$$\theta = \frac{\dot{A}}{A} + \frac{\dot{B}}{B} + \frac{\dot{C}}{C}. \qquad (4)$$

The overhead dots on the metric potentials represent the ordinary time derivatives. Defining the directional Hubble parameters in the X-, Y- and Z-axes as $H_1 = \frac{\dot{A}}{A}$, $H_2 = \frac{\dot{B}}{B}$ and $H_3 = \frac{\dot{C}}{C}$ respectively, the mean Hubble parameter can be written as $H = \frac{1}{3}(H_1 + H_2 + H_3)$ and $\theta = u^l_{;l} = 3H$. The field equation (1), for the metric (3) now assumes the explicit forms

$$\frac{\ddot{B}}{B} + \frac{\ddot{C}}{C} + \frac{\dot{B}\dot{C}}{BC} = -8\pi G \bar{p} + \Lambda \qquad (5)$$

$$\frac{\ddot{A}}{A} + \frac{\ddot{C}}{C} + \frac{\dot{A}\dot{C}}{AC} = -8\pi G \bar{p} + \Lambda \qquad (6)$$

$$\frac{\ddot{A}}{A} + \frac{\ddot{B}}{B} + \frac{\dot{A}\dot{B}}{AB} - \frac{h^2}{A^2} = -8\pi G \bar{p} + \Lambda \qquad (7)$$

$$\frac{\dot{A}\dot{B}}{AB} + \frac{\dot{B}\dot{C}}{BC} + \frac{\dot{A}\dot{C}}{AC} - \frac{h^2}{A^2} = 8\pi G \rho + \Lambda \qquad (8)$$

$$h\left(\frac{\dot{B}}{B} - \frac{\dot{A}}{A}\right) = 0 \qquad (9)$$

It is evident from (9), that for $h \neq 0$, $\frac{\dot{B}}{B} = \frac{\dot{A}}{A}$ or $H_2 = H_1$ which implies that

$$B = \beta A \qquad (10)$$

where $\beta \neq 0$ is an integration constant. Along with the conservation law $T^{ij}_{;j} = 0$, the vanishing covariant divergence of the Einstein tensor $R_{ij} - \frac{1}{2} g_{ij} R$ yields,

$$\frac{\dot{G}}{G} = -\frac{\dot{\Lambda}}{8\pi G\rho},  \tag{11}$$

which suggests that, the time variation of Newtonian Gravitational constant is very much associated with that of the cosmological constant. It is necessary to consider a rolling down cosmological constant in classical cosmological models to account for the present accelerated expansion of the universe and hence it is essential to think of a time varying Newtonian Gravitational constant in the field equations.

The shear scalar $\sigma$ for the metric (3) is defined as

$$\sigma^2 = \frac{1}{2}\left[\sum_i H_i^2 - \frac{1}{3}\theta^2\right] = \frac{1}{3}(H_1 - H_3)^2  \tag{12}$$

The shear scalar may be taken to be proportional to the expansion scalar which envisages a linear relationship between the Hubble parameters $H_1$ and $H_3$,

$$H_3 = kH_1  \tag{13}$$

which leads to a relation between the metric potentials $C$ and $A$ as

$$C = A^k,  \tag{14}$$

where $k$ is a constant and is usually assumed to be positive. $k$ takes care of the anisotropic nature of the model. The expansion scalar and the shear scalar now be expressed in terms of the single time varying parameter $H_1$ as

$$\theta = (k+2)H_1  \tag{15}$$

and

$$\sigma^2 = \frac{1}{3}(1-k)^2 H_1^2.  \tag{16}$$

In our earlier works [34, 35], we have emphasized that a barotropic bulk viscous pressure defined through $\xi\theta = \zeta\rho$ with $\zeta \geq 0$ is required to explain the observed accelerated expansion of the universe. For a barotropic cosmic fluid the proper pressure is related to the energy density as $P = \omega_0\rho$, $0 \leq \omega_0 \leq 1$. The combined effect of the proper pressure and the bulk viscous barotropic pressure leads to a total negative effective pressure $\bar{p} = \omega\rho$, where the new equation of state parameter $\omega$ is related to the old one $\omega_0$ as $\omega_0 - \zeta$. The antigravity effect of the total negative effective pressure provides the necessary acceleration. The observational limits on the equation of state parameter $\omega$ from SN Ia data are $-1.67 < \omega < -0.62$ [41] and that from a combination of SN Ia data with CMB anisotropy and galaxy clustering statistics are $-1.3 < \omega < -0.79$ [42].

For a bulk viscous barotropic bulk viscous cosmic fluid, the field equations (5-8) reduce to

$$(k+1)\frac{\ddot{A}}{A} + k^2\left(\frac{\dot{A}}{A}\right)^2 = -8\pi G\omega\rho + \Lambda,  \tag{17}$$

$$2\frac{\ddot{A}}{A} + \left(\frac{\dot{A}}{A}\right)^2 - \frac{h^2}{A^2} = -8\pi G\omega\rho + \Lambda \ , \tag{18}$$

$$(2k+1)\left(\frac{\dot{A}}{A}\right)^2 - \frac{h^2}{A^2} = 8\pi G\rho + \Lambda \ . \tag{19}$$

The conservation law $T^{ij}_{;j} = 0$ yields

$$H_1 = -\frac{1}{(\omega+1)(k+2)}\frac{\dot{\rho}}{\rho} \tag{20}$$

which after integration gives

$$A = \frac{k_2}{\rho^{1/\{(\omega+1)(k+2)\}}} , \tag{21}$$

where, $k_2$ is a positive integration constant. From eqns (17) and (18) we get,

$$\frac{\ddot{A}}{A} + (k+1)\left(\frac{\dot{A}}{A}\right)^2 = \frac{h^2}{(1-k)} \tag{22}$$

which admits the solution

$$A = h_1 t + k_3 \tag{23}$$

where $k_3$ is an integration constant which simply brings a shift in the time scale and hence can be neglected. $h_1 = \frac{h^2}{(1-k^2)}$ is a newly defined constant depending on the exponent $h$ and the anisotropic parameter $k$. The metric potentials can be expressed as

$$A = h_1 t \ , \ B = \beta h_1 t \ \text{and} \ C = (h_1 t)^k \tag{24}$$

and the line element(3) becomes

$$ds^2 = -dt^2 + (h_1 t)^2 \, dx^2 + (\beta h_1 t)^2 \, e^{-2hx} \, dy^2 + (h_1 t)^{2k} dz^2 \ . \tag{25}$$

From (21) and (24), we get,

$$\rho = \frac{\rho_0}{t^{\{(k+2)(\omega+1)\}}} \tag{26}$$

and consequently, the proper pressure

$$p = \frac{\omega_0 \rho_0}{t^{\{(k+2)(\omega+1)\}}} \tag{27}$$

where, $\rho_0 = \left(\frac{k_2}{h_1}\right)^{\{(k+2)(\omega+1)\}}$. The time variation of the rest energy density and the proper pressure depends on the value of the equation of state parameter $\omega$. If $\omega < -1$, then the rest energy density and proper pressure increase with the growth of time and the other way around if $\omega > -1$. Since it is logical to presume that the rest energy density of the universe should decrease with the growth of cosmic time, the reasonable limit for the equation of state parameter should be

$\omega > -1$. The mean Hubble parameter, scalar expansion, the shear scalar and the coefficient of bulk viscosity can be expressed as

$$H = \left(\frac{k+2}{3}\right)\frac{1}{t}, \tag{28}$$

$$\theta = (k+2)\frac{1}{t}, \tag{29}$$

$$\sigma = \frac{1-k}{\sqrt{3}}\frac{1}{t} \tag{30}$$

and $$\xi = \left(\frac{\zeta\rho_0}{k+2}\right)\frac{1}{t^{\{(k+2)(\omega+1)-1\}}}. \tag{31}$$

In the beginning of universe, i.e. at $t \to 0$, the Hubble parameter, the scalar expansion and the shear scalar assume infinitely large values where as with the growth of cosmic time they decrease to null values as $t \to \infty$. Since $\lim_{t\to\infty}\frac{\sigma}{\theta} \neq 0$, the anisotropy of the model is maintained throughout. The coefficient of viscosity $\xi$ evolves with time and its evolution depends on $k$ and $\omega$. If $\omega = \frac{1}{k+2} - 1$, $\xi$ remains a constant throughout the evolution of the universe. For $\omega > \frac{1}{k+2} - 1$, it decreases with time whereas it increases for $\omega < \frac{1}{k+2} - 1$.

The volume scale factor of the universe is

$$\tau = \beta(h_1 t)^{(k+2)} \tag{32}$$

and the radius scale factor can be $a = \beta^{1/3} h_1^{\left(\frac{k+2}{3}\right)} t^{\left(\frac{k+2}{3}\right)}$. The volume scale factor of the universe increases with the growth of cosmic time. The expansion or contraction of the volume scale factor depends on the parameter k. The deceleration parameter $q = -\frac{\tau\ddot{\tau}}{\dot{\tau}^2}$ gives us the nature of the expansion of the universe. If $q > 0$, the universe decelerates whereas if $q < 0$, then the universe accelerates. For the present model, the deceleration parameter is $q = \frac{1}{k+2} - 1$. For a given choice of the anisotropic parameter $k$, the deceleration parameter is a constant quantity and does not depend on the growth of cosmic time. Since $k$ is presumed to be a positive quantity, the deceleration parameter $q$ is negative i.e. $q < 0$ for all possible values of $k$, implying an accelerated expansion of the universe.

The differentiation of (19) with respect to time along with (11), (24) and (26), results into

$$G = \frac{2k}{8\pi\rho_0(\omega+1)} t^{\{(k+2)(\omega+1)-2\}} \tag{33}$$

and consequently, $$\Lambda = \left(k(k+2) - \frac{2k}{\omega+1}\right)\frac{1}{t^2}. \tag{34}$$

The Newtonian gravitational constant $G$ varies with time. $G$ increases with the increase in time if $k > -\frac{2\omega}{\omega+1}$ and $G$ decreases with the increase in time if $k < -\frac{2\omega}{\omega+1}$. $G$ comes out to be a pure constant if $k = -\frac{2\omega}{\omega+1}$. The sign of $G$ depends on the effective equation of state parameter $\omega$. If

$\omega < -1$, $G$ will be negative and if $\omega > -1$, then $G$ will be positive. However, since a decaying rest energy density of the universe requires that $\omega > -1$, the sign of $G$ is positive. The cosmological constant $\Lambda$ evolves from a large value at the beginning of the universe to a very small value at a large cosmic time. Present day observation requires a small but positive cosmological constant in classical cosmological models to account for the observed accelerated expansion of the universe. It is evident from (34) that, $\Lambda$ can be positive only if $k > -\frac{2\omega}{\omega+1}$. The consideration of positive $\Lambda$ leads to an increasing $G$ through the cosmic evolution. It is interesting to note from the outcome of this model that, at the beginning of the universe, $G$ has a negligible value. As is evident from the field equation (1), the vanishing $G$ implies that, at the beginning of the universe, the dominant role for energy is played by the cosmological constant. From (33), it can be inferred that $\frac{\dot{G}}{G} = \frac{(k+2)(\omega+1)-2}{t}$ which along with the above consideration of $k > -\frac{2\omega}{\omega+1}$ comes out to be a positive quantity.

The relationship between the cosmological constant and the mean Hubble parameter becomes $\Lambda \propto H^2$. Such relationship has already been used earlier by Arbaab [7]. For a Machian cosmological solution, the quantity $G\rho$ should satisfy the condition $G\rho \propto H^2$. The BIII model satisfies this condition for Machian solution.

The critical energy density $\rho_c = \frac{3H^2}{8\pi G}$, and the critical vacuum energy density $\rho_c = \frac{\Lambda}{8\pi G}$ for the anisotropic BIII model can be expressed, respectively, as

$$\rho_c = \frac{(k+2)^2(\omega+1)}{2k} \rho ,  \tag{35}$$

$$\rho_\Lambda = \frac{\{(k+2)(\omega+1)-2\}}{2} \rho .  \tag{36}$$

Consequently, the mass density parameter $\Omega_m = \frac{\rho}{\rho_c}$ and the density parameter corresponding to the vacuum $\Omega_\Lambda = \frac{\rho_\Lambda}{\rho_c}$ can be,

$$\Omega_m = \frac{2k}{(k+2)^2(\omega+1)}  \tag{37}$$

and $$\Omega_\Lambda = \frac{2}{\{(k+2)(\omega+1)-2\}}  \tag{38}$$

respectively.

In summary, we have investigated bulk viscous anisotropic cosmological model in the framework of General Relativity with variable gravitational and cosmological constants. The time varying nature of the cosmological constant is very much linked to the time variation of the gravitational constant. In order to get viable models, we have assumed that the shear scalar is proportional to the scalar expansion and the contribution of bulk viscosity to pressure is proportional to the rest energy density. Bulk viscosity is necessary to explain the accelerated expansion of the universe. The bulk viscous pressure along with the proper pressure results in a total effective negative pressure which provides the necessary antigravity effect to drive the acceleration.

The time varying nature of $\Lambda$ and $G$ depends on the bulk viscosity and the anisotropy of the universe. However, for small but positive cosmological constant at a later epoch, the Newtonian gravitational constant increases with the increase in cosmic time. The extracted time variation of the $G$ satisifes the Machian cosmological solution. The cosmological constant rolls down from a very large value at the beginning of the universe to a small positive value. The cosmological models considered along with the observational bounds restrict the value of effective equation of state parameter $\omega$ within the range $\omega > -1$ which is close to the observational limits of the combined SN Ia data with the CMB anisotropy and galaxy clustering statistics.